\documentstyle[epsfig,aps,preprint,floats]{revtex}
\begin{document}

\title{Dynamics of Stripes in Doped Antiferromagnets}
\author{C. Morais Smith$^a$, Y. Dimashko$^a$, N. Hasselmann$^{a,b}$ 
 and A. O. Caldeira$^c$ }
\address{$^{a \,}$I Institut f{\"u}r Theoretische Physik, Universit{\"a}t 
Hamburg, D-20355 Hamburg, Germany \\ 
$^{b \,}$Dept. of Physics, University of California, Riverside, CA, 92521, 
USA\\
$^{c \,}$Instituto de F{\'\i}sica Gleb Wataghin, Universidade Estadual de 
Campinas, CP 6165, 13085-970 Campinas SP, Brasil }

\date{\today}
\maketitle
\widetext
\vspace*{-1.0truecm}
\begin{abstract}
\begin{center}
\parbox{14cm}{
\noindent
We study the dynamics of the striped phase, which has previously been suggested to be the
ground state of a doped antiferromagnet. Starting from the $t-J$ model, we derive
the classical equation governing the motion of the charged wall by using 
a ficticious spin model as an intermediate step. A wave-like equation of motion is 
obtained and the wall elasticity and mass density constants are derived in terms 
of the $t$ and $J$ parameters.
The wall is then regarded as an elastic string which will be trapped by the
pinning potential produced by randomly distributed impurities. We evaluate
the pinning potential and estimate the threshold electric field which has to
be applied to the system in order to release the walls. 
Besides, the dynamics of the stripe in the presence of a bias field below the 
threshold is considered and the high- and low-temperature relaxation rates
are derived.  
}
\end{center}
\end{abstract}

\narrowtext

The discovery of a deep connection between superconductivity and quantum 
antiferromagnetism in the phase diagram of the cuprate perovskites has 
stimulated various attempts to understand the effects of dilute holes in a 
spin 1/2 Heisenberg antiferromagnet. The problem has usually been addressed 
by assuming that a doped antiferromagnet can be described by
a gas of holes with uniform density. However, several calculations
\cite{Schu,Poil-Rice,Inui-Litt,Zaan-Gunn,Pokr-Kala,Giam,Prel-Zoto}
suggest that in this system there is a modulation of the charge and
spin densities, i.e., the holes cluster along lines which
separate undoped antiferromagnetic domains (striped phase).

Experimentally, recent measurements also indicate that striped order indeed 
occurs in doped planar antiferromagnets. In the insulating nickel-oxides, such
stripe modulations were reported \cite{Tran-Butt,Sach-Butt,Tran-Lore} and
the data were consistent with multiband Hubbard model calculations 
\cite{Zaan-Litt}. 
For the case of the copper-oxides, elastic neutron diffraction experiments
have revealed static stripe order in the non-superconducting 
compound $\rm La_{1.48}Nd_{0.4}Sr_{0.12}CuO_4$ \cite{Tran-Ster} while in 
the superconducting compound $\rm La_{2-x}Sr_{x}CuO_4$ inelastic scattering peaks
at incommensurate wave vectors suggest the existence of a very similar, albeit 
slowly fluctuating, striped phase \cite{aeppli,Yamada}.
Although incommensurate spin fluctuations are not observed in the low
doping region of the cuprates, muon spin resonance and nuclear 
quadrupole resonance experiments on $\rm La_{2-x}Sr_{x}CuO_4$ with 
$0 \le x \le 0.018$ \cite{Bors} have been successfully interpreted within 
models that presume a striped structure \cite{antonio}.
The surprisingly strong suppression of superconductivity in these materials 
for even low Zn doping was also found to be consistent with the existence of
stripes \cite{antonio2}.

The objective of the present work is to study the dynamics of this striped 
phase in the low doping regime, where interactions between neighbouring stripes
is assumed to be negligible.
We treat the problem on the basis of the $t-J$ model and
establish the connection to the discrete elastic theory. 
The domain wall is then considered from a phenomenological point of 
view, i.e., as an elastic line trapped by the pinning potential
produced by impurities. The depinning of the line from the potential well by
applying an electrical field perpendicular to the stripe structure is
investigated, and the threshold field corresponding to the onset of a 
state with mobile lines is determined. Finally, the relaxation process in the
presence of a bias field below the threshold is considered and the classical and
quantum decay rates from the metastable state are computed. 
 
The first calculations of the striped phase in 2D antiferromagnets with holes
have been done within the Hubbard model in the vicinity of half filling.
They were based on the self-consistent Hartree-Fock formalism and have been
performed for small and large values of the ratio $U/t$. The results can
be summarized as follows: In the small $U$-approximation, vertical
domain walls (parallel to the $x$- or $y$-axis) are stable 
\cite{Schu,Inui-Litt}, whereas for large $U$ diagonal walls are 
energetically more favorable \cite{Poil-Rice,Inui-Litt}.  
This crossover from vertical to diagonal stripes was numerically
calculated to happen at $U / t \sim 3.6$ \cite{Inui-Litt} (see Fig.\ 1).

In the $U/t \gg 1$ limit, the Hubbard model with almost half-filled band can
be reduced to the $t-J$ model with effective exchange constant $J = 4 t^2/U$
\cite{Rice}. Contrary to the Hubbard model, where each site of the lattice
corresponds to four possible states, in the $t-J$ model only three states are
allowed, since double occupancy is forbidden. Hence, the dimensionality of
the Hilbert space in the latter model is much smaller and for finite clusters
an exact diagonalization of the system, without using the Hartree-Fock 
approximation, is possible. These studies of the $t-J$ model have been
recently performed and they confirm the results obtained previously: by exactly
diagonalizing small systems, Prelovsek and Zotos \cite{Prel-Zoto} have 
verified that a striped phase with holes forming domain walls along the (0,1)
or (1,0) direction arises for $J > J_s \sim 1.5 t$, while in the regime
$0.4 < J/t < 1.2$ the domain walls appear along the (1,1) direction. Besides,
they found signs of phase separation into a hole-rich and a hole-free phase
at $J > J_s^* \sim 2.5 t$. In Fig.\ 1 we grafically represent both, the results 
obtained from the $t-J$ and from the Hubbard model in order to make their comparison
clearer. 

The obtained results for the striped phase can be understood on simple 
physical grounds. In Figs.\ 2 
(a) and (b), diagonal and vertical stripes (ensemble of holes) embbeded in 
a AF background are respectively represented. The holes can jump from
one site to another and during this process they gain kinetic energy $t$. An
inspection of Fig.\ 2 leads us to conclude that the diagonal 
configuration favours the dynamics (in this case the holes can move in 2D,
horizontally or vertically), whereas for the vertical stripe the holes
are confined to move only in the horizontal direction. On the other hand,
in the vertical configuration the holes are closer to each other and hence
there is a gain in the segregation energy $J$. Then the final behavior results
from the competition between the kinetic $(t)$ and the segregation $(J)$ 
energies (see Fig.\ 2): When $t > J = 4 t^2 /U$, i.e., for $U / t$ large, the 
kinetic energy dominates over the segregation energy and the stripes are 
diagonally arranged. When the gain in the segregation energy becomes more 
relevant, for $t < J $, the vertical (horizontal) formation arises. 

Recently, the stripe dynamics has been studied from a different perspective:
the domain walls were regarded from a phenomenological point of view as
vibrating strings \cite{Zaan-Horb-Saar}.
The motion of a single stripe was initially considered, and the
analysis was then extended to the more complicated regime involving many
(interacting) domain walls. Here, we adopt a similar description. However,
instead of using the initial assumption that the stripe can be described
by an elastic line, we start from the $t-J$ model and {\it show} that the
classical equation of motion describing the stripe dynamics in the limit of
long wavelength displacements has a wave-like form. In this way, we deduce
the phenomenological mass and elastic coefficients for the string in 
connection with the ``microscopic'' $t$ and $J$ parameters. For the sake
of simplicity, we concentrate on the vertical configuration. This implies
that our considerations hold for $J_s^* > J > J_s$, i.e., $2.5t > J > 1.5 t$
\cite{Prel-Zoto}. Besides, we study the problem in the dilute 
limit, when the doping concentration is low and we can investigate the 
behavior of a single stripe (chain of holes). 

In order to derive the quantum equation of motion describing the dynamics of 
the $n$-th hole in the chain, we map the initial $t-J$ Hamiltonian onto a quantum
spin chain problem with large spin. 
Replacing the spin-chain operators in the quantum equation 
of motion by their classical values and considering the long wavelength limit, 
a classical wave-type equation of motion is derived for the stripe. The
justification for this classical approach rests on the assumption of the existence
of zero mode excitations, i.e., we assume that a continuum description of the
problem is possible and that the discreteness of the original lattice formulation
is irrelevant in the long wavelength limit. Recently, it was shown \cite{Hassel}
that the striped phase undergoes a roughening transition, i.e., the flat 
(gapped) phase present at higher values of $J/t$ becomes rough (gapless) as
$J/t$ is reduced. Hence, our approach applies for the rough phase of the stripe, 
in which the underlying lattice structure is unimportant
and the behaviour of the system is governed by a gaussian fixed point. 

Since the equation of motion describing the stripe dynamics has a wave-like
form, the stripe is from then onwards regarded as an elastic string. 
The random pinning potential due to the presence of impurities is
evaluated and the activation energy barrier is calculated for the case of
a bias electrical field applied perpendicularly to the stripe, both in the
high temperature phase, where quantum fluctuations can be neglected, and
in the quantum dominated low temperature phase. The threshold field is estimated 
and we briefly discuss if and how stripe depinning could be experimentally 
observed. Besides, the dissipation effects are considered and their
influence in the stripe depinning is discussed. 

The outline of the paper is as follows: in section I, the model is
introduced and the equation of motion is derived. The pinning potential 
and the threshold field are calculated in section II.
In section III we investigate the classical and quantum relaxation regimes 
at fields below the threshold. The discussions and conclusions are presented 
in section IV. 

\section{THE MODEL}

The $t-J$ Hamiltonian describing a chain of holes (stripe) embbeded in an
antiferromagnetic background is \cite{Rice}
\begin{equation}
H_{tJ} = -t \sum_{<ij>} c^{\dagger}_{i \sigma} c_{j \sigma} + J
\sum_{<ij>} \left(\vec{S}_i \cdot \vec{S}_j - \frac{1}{4} n_i n_j \right).
\end{equation}
Here, $H_{tJ}$ acts in the truncated Hilbert space consisting of states for
which the double occupancy of any site is forbidden, i.e., $n_i=0,1.$

We apply this Hamiltonian for investigating the dynamics of an infinitely long 
stripe. The linear concentration of holes in the stripe is assumed to be one 
hole per one lattice constant, so we can neglect charge and spin fluctuations
within the stripe. The stripe represents a domain boundary (DB) dividing the 
antiferromagnetic plane into two N\'eel's phases. We consider these two phases
to have opposite staggered magnetization ({\it antiphase DB}). Hence, the 
string can move without disturbing the initial magnetic order, i.e., it is 
delocalized (see Fig. 3a). This case is opposite to the other hypothetical 
case, when these two staggered fields are equivalent ({\it phase DB}) 
and the string is localized, since its motion is strongly frustrated by the
surrounding magnetic order (Fig. 3b). 

Here, we concentrate on the dynamics of the vertical configuration and
assume that the stripe is located along the $y$-direction. Further, we make one
additional hypothesis, which reduces the problem to a directed, but discrete,
polymer problem: the 
holes are constrained to move along the transversal $x$-direction only, i.e.,
we neglect overhangs. 
This assumption is true for the shortest excursions of the holes from the 
initial vertical position of the string (up to one step) and therefore seems
to be plausible as far as one considers only small magnitudes of the oscillations.
Hence, the supposition holds for the $J > t$ case, when larger excursions of
the holes are suppressed by the dominating exchange $J-$term. In the opposite
$J < t$ case, the hypothesis ceases to be valid, and the one-dimensional
approximation cannot be used anymore. This corresponds to the more complicated 
case of the diagonal DB, which will be treated in a separate work.  

Let us denote $u_n$ the displacement along the $x-$direction of the $n-$th hole from
its equilibrium position. We describe the relative displacement of two 
neighbouring holes by $\sigma _n^z  = u_n-u_{n-1}$ and treat this value as the 
 $z-$component of some effective local spin $\vec{\sigma}_n$. This spin is related 
to the $n-$th segment of the stripe. Displacement of the $n-$th hole by one step 
along the positive direction of the $x-$axis increases 
 $\sigma_n^z (\sigma_n^z \to \sigma_n^z + 1)$ and decreases 
 $\sigma_{n+1}^z (\sigma_{n+1}^z \to \sigma_{n+1}^z -1)$. 
Their sum remains unchanged. Hence, the motion of the $n-$th hole is
described by the spin operator $-t \sigma^{-2} (\sigma_{n+1}^+ \sigma_n^- +
\sigma_{n+1}^- \sigma_n^+)$. The $\sigma^{-2}$ factor provides the 
conservation of the norm under the action of this term on the spin-wave
function.
Besides, by accounting for the different N\'eel order on the left and right
hand sides of the stripe, we see that any increase of the relative 
displacement of two neighbouring holes by unity results in a increase of the
exchange energy by $J/2$. Thus, the contribution of the $n-$th
segment to the exchange energy of the string is $(J/2) \mid \sigma_n^z \mid$. 

A similar problem was investigated by Eskes et al. \cite{Eske-et},
but the Hilbert space in their work was restricted in a different way: only
configurations where two neighbouring holes are separated by one lattice unity
were allowed. Hence, they mapped the relative displacement coordinate onto 
a spin one chain and determined the phase diagram features, in the same 
spirit as in Refs. \cite{denN,Schu1}.
Here, we consider a broader Hilbert space and search for the classical 
equations of motion instead of the quantum phase-diagram. 

The projection of the initial $t-J$ Hamiltonian onto the new Hilbert space 
gives rise to the spin-chain problem with the effective Hamiltonian
\begin{equation}
H=-\frac{t}{\sigma^2}
\sum_n(\sigma _{n+1}^{+}\sigma _n^{-}+\sigma _{n+1}^{-}\sigma
_n^{+})+\frac{J}{2}\sum_n \mid \sigma _n^z \mid 
\end{equation}
and the standard commutation relations
$$
[\sigma _n^z,\sigma _m^{\pm }]=\pm \delta _{n,m}\sigma _m^{\pm },\qquad
[\sigma _n^{+},\sigma _m^{-}]= 2 \delta _{n,m}\sigma _m^z. 
$$
The effective spin $\sigma$ is the maximal relative displacement of two
neighbouring holes.

The configuration of the stripe is determined by the $z-$projection of the spin,
 $\sigma_n^z$. The quantum equation governing the motion of the $\sigma_n^z$ operator
is
\begin{equation}
\hbar^2 \stackrel{..}{\sigma}_n^z = - \left[ \left[\sigma_n^z,H\right],H\right].
\label{sigmanz}
\end{equation}
The commutators involving the $J-$term of the Hamiltonian (which contains the
 $\mid \sigma_n^z\mid$) can be more easily evaluated with the help of the general
relation
\begin{equation}
[\sigma_n^\pm, f(\sigma_m^z)] = \sigma_n^\pm
\Big( f (\sigma_m^z) - f (\sigma_m^z \pm 
\delta_{n,m}) \Big). 
\label{com}
\end{equation}
Here, $f(\sigma_m^z)$ is an arbitrary function of the operator $\sigma_m^z$. 

The next step is to evaluate the commutators on the right-hand side of Eq. 
(\ref{sigmanz}). Then, we consider the classical limit $\sigma \to \infty$ and
replace the spin-operators $\sigma_n^z$ and $\sigma_n^{\pm}$ by the c-numbers:
\begin{equation}
\sigma_n^z \to \sigma \sin \alpha_n, \qquad \sigma_n^\pm \to \sigma \exp (\pm i \phi_n
) \cos \alpha_n.
\end{equation}
In the long wavelength limit, the discrete variable $n$ can be replaced by a
continuous one $y$. Hence, 
\begin{equation}
\mid \phi_{n+1} - \phi_n \mid = a \left| \frac{\partial \phi}{\partial y} \right|
 \ll 1, \qquad 
\mid \alpha_n \mid = a\left| \frac{1}{\sigma}\frac{\partial u}{\partial y} \right|
\ll 1
\end{equation}
and the equation of motion (\ref{sigmanz}) acquires the simple wave-like form
\begin{equation}
\left(\frac{\partial^2}{\partial t'^2} - \frac{t Ja^2}{\hbar^2} 
\frac{\partial^2}{\partial y^2} \right) \alpha = 0.
\label{eqmtheta}
\end{equation}
Here, $t'$ denotes the time and $a$ is the lattice spacing. 
A more detailed derivation is given in the Appendix A.

Due to the differential relation between the $\alpha$ and $u$ variables, $\alpha =
(a/\sigma) (\partial u / \partial y)$, the displacement of stripe obeys the same
wave-like equation,
\begin{equation}
\left(\frac{\partial^2}{\partial t'^2} - \frac{t Ja^2}{\hbar^2} 
\frac{\partial^2}{\partial y^2} \right) u = 0.
\label{imp}
\end{equation}
This equation describes the long wavelength oscillations of the stripe around the 
equilibrium position $u = const.$ The corresponding action reads
\begin{equation}
{\cal S} [u(y,t')] = \int_{-\infty}^{\infty}\, dy \int_{0}^{\infty}\, dt'
\left[ \frac{\hbar^2}{2 ta^3}
\left(\frac{\partial u}{\partial t'}\right)^2 - \frac{J}{2a}
\left(\frac{\partial u}{\partial y}\right)^2 \right]
\label{elasticaction}
\end{equation}
The problem then has been mapped onto a massive string with linear mass 
density $\rho = \hbar^2/ta^3$ and elastic tension coefficient $C = 
J/a$. 

Finally, one can observe from Eq. (\ref{imp}) that the long wavelength 
elementary excitations of the string are gapless and have
the phonon-like dispersion relation
\begin{equation}
\omega = c k,
\end{equation}
with phase velocity $c = a \sqrt{t J}/\hbar$.

We emphasize again that our derivation assumes the validity of a continuum
description. Hence it holds for the rough phase. 
 
\section{THRESHOLD FIELD}

Up to now, we have only considered the dynamics of a stripe embedded in an
antiferromagnetic background. However, in real systems, we have also to take 
into account the presence of randomly distributed point-like impurities which
act to pin the lines, leading to a glassy phase with trapped stripes.
In order to reduce the pinning barrier, we apply an external electrical field
perpendicularly to the stripes formation. Then, we determine the threshold 
field above which the potential barrier vanishes and the stripes can flow
through the sample, as in a liquid state. 

The free energy describing
an elastic string along the $y$-direction, which tends to move due to the
action of an externally applied electrical field $E$ competing against the
pinning barrier $V_{pin}$ is
\begin{equation}
{\cal F} [u(y)] = \int_{-\infty}^{\infty}\, dy \left[
\frac{C}{2} \left(\frac{\partial u}{\partial y}\right)^2 +
V_{pin}  - \frac{e E u}{a} \right]. \label{freeen}
\end{equation}

The next step now is the evaluation of the pinning potential $V_{pin}$.
Let us consider that the pinning mechanism is produced by the ionized acceptors
which sit on a plane parallel to and close to
the CuO-plane. An impurity with two dimensional coordinates $\vec{R}$
produces at the position $\vec{r}$ in the CuO-plane
the Coulomb potential $G(\vec{R}-\vec{r})$
\begin{equation}
G(\vec{R}-\vec{r})= \frac{e^2}{\epsilon |\vec{R} - \vec{r}|}  
\end{equation}
The total potential felt at position $\vec{r}$ in the CuO planes can
then be written as
\begin{equation}
V_{pin}(\vec{r})=\frac{1}{a^2}\int d^2 \vec{R} {\cal N}(\vec{R}) 
G(\vec{R}-\vec{r}),
\end{equation}
where ${\cal N}(\vec{R})$ is the number of impurities (zero or one) 
at the position $\vec{R}$. Hence, 
${\cal N}(\vec{R})=\left({\cal N}(\vec{R})\right)^2 $. Let us denote the average 
number of impurities per one site $\left< {\cal N}(\vec{R})\right >=\nu$, where 
 $\left< \dots \right>$ represents average over the disordered impurity ensemble 
and $0<\nu<1$. If the impurity distribution would be uncorrelated, 
we could write the density-density correlator as
\begin{equation}
\rho _{0}=\left< {\cal N}(\vec{R}){\cal N}(\vec{R}')\right >
-\left< {\cal N}(\vec{R})\right >\left< {\cal N}(\vec{R}')\right >
=\nu (1-\nu )\delta (\vec{R}-\vec{R}').
\end{equation}
Its Fourier-transform is a constant for any value of the wave vector $\vec{k}$,
$\rho_{0}(\vec{k})=\nu (1-\nu)=const$. This correlator describes long-wave 
fluctuations with any $k$, including the small ones. Although the small-$k$
fluctuations are allowed by the statistics of the completely disordered state, 
in reality they are strongly suppressed by the long-range Coulomb interaction 
of the impurities, since such fluctuations provide large value of the Coulomb
energy $\sim k^{-1}$. Hence, the real correlator should be suppressed for small
$k$ at some scale $k<\lambda $. This cut-off can be introduced by the following 
choice of the correlation function:
\begin{equation}
\rho (\vec{k})=\frac{\nu (1-\nu)}{1+\lambda ^{2}/k^{2}}.
\end{equation}
Since the only length scale in the system of the impurities is the average 
distance between two neighbor impurities $u=a\nu ^{-1/2}$, the $\lambda$ 
parameter should be estimated as $\lambda \sim 1/u$.
Further, the two-point correlator of the impurity potential
is given by
\begin{eqnarray} \nonumber
{\cal K}(\vec{r}) & = &
\left<V_{pin}(\vec{r})V_{pin}(0)\right>_d - \left<V_{pin}(\vec{r})\right>_d 
\left<V_{pin}(0)\right>_d \\ 
\nonumber
& = & \frac{1}{a^4}
\int d^2 \vec{R} \ d^2 \vec{R}' G(\vec{R}') G(\vec{R} - \vec{r}) 
\rho(\vec{R},\vec{R}').
\end{eqnarray}
The Fourier transform of ${\cal K}(\vec{r})$ reads
\begin{equation}
{\cal K}(\vec{k})=G^{2}(\vec{k})\rho(\vec{k})
=(\nu-\nu^2) 
\left(\frac{e^2}{\epsilon a}\right)^2  \frac{4 \pi^2}{(k^2+\lambda^2)a^2},
\end{equation}
where $G(\vec{k})=2\pi e^{2}/(\epsilon k)$ is the 2D Fourier-transform of the 
Coulomb potential $G(\vec {r})$. 
Rewriting then the correlator of the potential in real space we find
\begin{equation}
{\cal K}(\vec{r})=2 \pi
\left(\frac{e^2}{\epsilon}\right)^2  (\nu-\nu^2) K_0(\lambda r), 
\end{equation}
where $ K_0(\lambda r)$ is the modified Bessel function.

Let us now define $\varepsilon_{pin} = \int dy V_{pin}$. 
If the stripe interacting with the random pinning potential is stiff, the
average pinning energy $\left< \varepsilon_{pin} (L) \right>_d$ of a segment
of length $L$ is zero. The fluctuations of the pinning energy, however,
remain finite,
\begin{equation}
\left< \varepsilon_{pin}^2 \right>_d = \frac{1}{a^2} \int_0^L dy dy' {\cal K}(0,y-y')
\simeq \gamma L
\label{energie}
\end{equation}
with 
\begin{equation}
\gamma= \frac{ 2 \pi^2 \varepsilon_c^2 \sqrt{\nu}}{a}. 
\label{gamma}
\end{equation}
Here, $\varepsilon_c = e^2 / \epsilon a$ denotes the Coulomb energy scale. 
The sublinear growth of $\left< \varepsilon_{pin}^2 (L) \right>_d^{1/2}$
is due to the competition between individual pinning centers. The dynamic
approach to this problem was introduced by Larkin and Ovchinnikov 
\cite{Lark} in the ``Collective Pinning Theory'' (CPT) for describing the 
dynamics of weakly pinned vortex-lines in the high temperature superconductors. 
A scaling approach was also considered in connection with the pinning
problem in charge-density-wave systems \cite{Fuku-Lee,Lee-Rice}. 
The results of the CPT can be summarized as follows \cite{Gianni}: 
Eq.\ (\ref{energie}) implies that a stiff stripe is never pinned,
since the pinning force grows only sublinearly, whereas the electrical
driving force increases linearly with length. On the other hand, due to
the elasticity, the stripe can acommodate to the potential on some
``collective pinning length'' $L_c$. Hence, each segment $L_c$ of the stripe
is pinned independently and the driving force is balanced. 

Our task now is to determine this length $L_c$. The evaluation of the 
free energy by using dimensional estimates provides
\begin{equation}
{\cal F} [u,L] \sim C \frac{u^2}{L} - \sqrt{\gamma L} - \frac{e E u L}{a}.
\label{freeenergie}
\end{equation}
By minimizing ${\cal F} [u,L] / L$ with respect to L at zero bias field
\cite{Gianni} we determine the collective pinning length $L_c$ along the 
string,
\begin{equation}
\frac{\delta {\cal F} / L}{\delta L} \Big|_{L_c} = 0, \qquad 
L_c\simeq  \left(\frac{C u^2}{\sqrt{\gamma}} 
\right)^{2/3}.
\end{equation}
Assuming that $u \sim \lambda^{-1} \sim a / \sqrt{\nu}$, the average impurity spacing, 
and using eq.\ (\ref{gamma}) we find 
\begin{equation}
L_c\simeq  a \nu^{-5/6} \left(\frac{ \varepsilon_l }{\varepsilon_c} 
\right)^{2/3},
\label{Lc}
\end{equation}
where the elastic energy $\varepsilon_l = C a.$
Experimentally, it is difficult to measure the collective length $L_c$. 
However, the threshold electric field $E_c$ corresponding to the vanishing of
the barrier is a quantity which can be easily
experimentally determined. The critical value $E_c$ can be 
estimated by equating the pinning energy $\sqrt{\gamma L_c}$ to the
electric energy $e E u L_c /a$. We then obtain
\begin{equation}
E_c = \frac{a}{eu}\, \sqrt{ \frac{\gamma}{L_c}} \sim \frac{ 
\nu^{7/6}\varepsilon_c }{ea} 
\, \left(\frac{\varepsilon_c}{\varepsilon_l}\right)^{1/3}. 
\end{equation}

For typical high-$T_c$ materials, such as La$_{2-\nu}$Sr$_\nu$CuO$_4$, $J \approx 
0.1$ eV, $a \approx 4$ \AA $\,$and $\epsilon = \epsilon_0 \approx 30$ \cite{Chen}. 
Hence, one can estimate the elastic energy $\varepsilon_l = J \approx 0.1$ eV and the
Coulomb energy $\varepsilon_c = e^2/(\epsilon a) \approx 0.1$ eV. For a doping 
concentration in the antiferromagnetic insulating phase, 
for instance, for $\nu = 10^{-3}$, we then obtain the collective pinning length 
 $L_c \approx 10^3$ \AA $\,$and the critical electrical field 
 $E_c \approx  10^3$ V/cm. 

\section{CLASSICAL AND QUANTUM RELAXATION PROCESS}

We have estimated the threshold field $E_c$ corresponding to the onset of the
motion of depinned stripes. Next, we are interested in studying the relaxation process
taking place at applied fields $E < E_c$. In this case there is a finite
pinning barrier preventing the motion of the stripe, but it can still jump
over (under) the barrier due to thermal (quantum) fluctuations. 

\subsection{Classical limit}

At high temperatures $T$, the decay rate
 $\Gamma_t$ is given by the Arrhenius law, $\Gamma_t \sim \exp(-U / T)$.
The activation energy $U$ can be determined by extremizing the free energy.
In order words, within the semiclassical approximation, the energy barrier
 $U$ is nothing but the free energy ${\cal F}$ evaluated at the saddle
point configuration $u_s$, $U = {\cal F} [u_s]$. As far as we are interested
in evaluating the decay rate only within exponential accuracy, we can
safely neglect the dynamical terms (as for instance, the kinetic one) in the
free energy, since these terms would give a correction only to the prefactor
multiplying the exponential function. 

By substituting the collective length $L_c$ as given by 
eq.\ (\ref{Lc}) into the free energy (\ref{freeenergie}), we can estimate 
the collective pinning energy barrier
\begin{equation}
U_c\simeq \sqrt{\gamma L_c}
\simeq \nu^{-1/6} (\varepsilon_l \varepsilon_c^2)^{1/3}.
\label{activ}
\end{equation}
For the case of La$_{2-\nu}$Sr$_\nu$CuO$_4$ with $\nu = 10^{-3}$ considered in the
previous section, we estimate the pinning barrier to be of the order 
of $10^3$ K. The barrier exhibits a weak dependence on the doping parameter,
 $U_c \propto \nu^{-1/6}$.
Notice that this barrier height is compatible with the estimates
presented in \cite{Zaan-Horb-Saar} for the binding energy of holes in a
domain wall. Besides, this value is also comparable to the one obtained for
the vortex creep process in high-$T_c$ superconductors, 
when $U_c \sim 10^2 - 10^3$ K. 

The dynamics of the stripe in the presence of the pinning potential is simply
an example of the more general problem of elastic manifolds in quenched 
random media. Investigations of the statistical mechanics of this object have
shown that a stripe confined to move in a plane is always in a pinned phase,
\begin{equation}
\left< \left<[u(L) - u(0)]^2 \right> \right> \, \sim \, u_c^2 \left( 
\frac{L}{L_c} \right)^{2 \zeta}, \qquad L > L_c,
\end{equation} 
with a wandering exponent $\zeta = 2/3$ \cite{Huse-Henl-Fish,Kard}. Here,
$\left< \left<...\right> \right>$ denotes the full statistical average 
over dynamical variables (thermal) and over disorder, $L$ is the distance 
along the stripe, and $u_c$ and $L_c$ are transverse and longitudinal 
scaling parameters, respectively.
In our case, $u_c \sim \lambda^{-1}$ (scale of the disorder potential) and
$L_c$ is the collective pinning length.

Besides, it was also found that competing metastable states that differ from
one another on a length $L$ are separated by a distance
\begin{equation}
u(L) \sim u_c \left(\frac{L}{L_c}\right)^\zeta, \qquad L > L_c
\end{equation}
and a typical energy barrier
\begin{equation}
{\cal U}(L) \sim U_c \left( \frac{L}{L_c} \right)^{2 \zeta - 1}, \qquad L > L_c,
\label{optbar}
\end{equation}
where $U_c$ denotes the scaling parameter for energy. For the single
stripe problem, $U_c$ reduces to the collective pinning energy. 
The free-energy functional at low driving fields $E \ll E_c$ is
\begin{equation}
{\cal F} (L) \sim U_c \left(\frac{L}{L_c} \right)^{2 \zeta - 1} -
\frac{e E L_c u_c}{a} \left( \frac{L}{L_c} \right)^{\zeta + 1}.
\label{optfe}
\end{equation}

The problem now has been reduced to a nucleation process \cite{Lang}.
If a nucleus with length $L$ larger than some optimal length $L_{opt}$
is formed, the system will move to the next minimum. On the other hand,
if the activated segment is smaller than the optimal one, the nucleus
will collpase to zero. The optimal nucleus can be found by extremizing
the free-energy, $\partial_L {\cal F} (L) |_{L = L_{opt}} = 0$ and we
obtain
\begin{equation}
L_{opt} (E) \sim L_c \left(\frac{E_c}{E}\right)^{1/(2-\zeta)} 
\label{optl}
\end{equation}
Inserting (\ref{optl}) back into the free energy (\ref{optfe}) we
verify that the minimal barrier for creep increases algebraically
for decreasing bias field, 
\begin{equation}
U(E) \sim U_c \left( \frac{E_c}{E} \right)^\mu
\label{mainone}
\end{equation}
with $\mu = (2 \zeta - 1 )/(2 - \zeta) = 1/4$. 
Hence, the system is in a glassy phase, with a diverging barrier 
in the limit of vanishingly small applied electrical fields.

Another interesting limit to study the dynamical behavior of the
stripe, is at fields below but close to the critical field $E_c$, 
i.e., at $E_c - E \ll E_c$. In this case, the effective potential 
given by the pinning and the bias electrical field terms can be 
written as \cite{Gianni}
\begin{equation}
V_{eff} (u) = V_F \left[ \left(\frac{u}{u_F}\right)^2 - 
\left(\frac{u}{u_F}\right)^3 \right]
\end{equation}
with $V_F \sim V_c (1 - E/E_c)^{3/2}$ and $u_F \sim u_c 
(1 - E/E_c)^{1/2}$.
The critical potential barrier $V_c = e E_c u_c /a$.
The energy of a distortion $u_F$ of the stripe on a scale $L_F$ 
is estimated to be
\begin{equation}
{\cal E}(u_F, L_F) \sim \left[ \frac{C}{2} \left(\frac{u_F}{L_F} 
\right)^2 + V_{eff} (u_F) \right] L_F.
\end{equation}

The competition between the barrier to be overcome $V_F$ with the
elastic energy density $C u_F^2 / L_F^2$ determines the length of
the saddle-point configuration
\begin{equation}
L_{FS} \sim u_c \sqrt{\frac{C}{V_c}} \left( 1 - \frac{E}{E_c} 
\right)^{-1/4}  \sim L_c \left( 1 - \frac{E}{E_c}
\right)^{-1/4}.
\end{equation}

Finally, the energy barrier for thermal activation of the stripe out
of the pinning potential reads
\begin{equation}
U(E) \sim u_c V_c \left( \frac{C}{V_c} \right)^{1/2} \left(1 - \frac{E}{E_c}
\right)^{5/4} \sim U_c \left(1 - \frac{E}{E_c}\right)^{5/4}.
\label{maintwo}
\end{equation}

Eqs. (\ref{mainone}) and (\ref{maintwo}), together with (\ref{activ}) are
the main results of this section. From an experimental point of view, it 
would be easier to observe the creep of the stripe near criticality, where
the thermal process is described by Eq. (\ref{maintwo}) and the activation
barrier (\ref{activ}) can be reduced by one or two orders of magnitude
due to the presence of the bias electrical field.

\subsection{Quantum limit}

At low temperatures, we expect the decay process to be driven by quantum
fluctuations. In this case, the dynamical terms become essential since they
are related to the so-called traversal time. The tunneling rate is then
 $\Gamma_q \sim \exp(-B/\hbar)$, where $B$ is given by the Euclidean action
 ${\cal S}_E$ of the system (the action in an imaginary time formalism)
evaluated at the saddle point solution, $B = {\cal S}_E [u_s]$. 
The total Euclidean action describing the elastic domain
wall in the presence of random impurities and an external electric field is
\begin{equation}
{\cal S}_E [u(y,\tau)] = \int_{-\infty}^{\infty}\, dy \int_{0}^{\infty}\, d\tau
\left[ \frac{\rho}{2}
\left(\frac{\partial u}{\partial \tau}\right)^2 + 
\frac{C}{2}\left(\frac{\partial u}{\partial y}\right)^2 
 + V_{pin}
- \frac{e E u}{a} \right].
\label{euclaction}
\end{equation}

The tunneling time $\tau_c$ can be estimated by equating the
kinetic and elastic terms, 
\begin{equation}
\rho \frac{u_c^2}{\tau_c^2} \sim C \frac{u_c^2}{L_c^2}.
\end{equation}
We then obtain $\tau_c \sim L_c \sqrt{\rho/C}$. Substituting
$\tau_c$ into the Euclidean action (\ref{euclaction}), we find
\begin{equation}
B_c^m \sim \tau_c L_c \frac{C u_c^2}{L_c^2} \sim u_c^2 \sqrt{\rho
C}  \sim \frac{\hbar}{\nu} \, \sqrt{\frac{\varepsilon_l}{t}}.
\label{bmas}
\end{equation}
It is important to notice that the extremal value of the action $B_c^m$ does
not depend on the collective pinning length $L_c$ and hence it is independent
of the pinning potential. 

Next, we account for dissipation effects in order to 
generalize our model. Since we cannot determine the friction coefficient 
 $\eta$ from microscopic calculations (we are considering frozen N\'eel phases for 
describing the antiferromagnetic background), we will evaluate it in a 
phenomenological way. Moreover, we restrict ourselves to the simplest case of 
ohmic dissipation and study the problem within the framework of the
Caldeira-Leggett model \cite{Cald-Legg}. 

In the overdamped limit, when $\eta \tau_c / \rho \ll 1$, we can neglect
the massive term in equation (\ref{euclaction}) and substitute 
\begin{equation}
\frac{\rho}{2} \left(\frac{\partial u}{\partial \tau}\right)^2 \qquad \to \qquad
\int_0^\infty d\tau' \frac{\eta_l}{4 \pi} 
\left[\frac{u(\tau) - u(\tau')}{\tau - \tau'} \right]^2.
\end{equation}
The tunneling time now can be obtained by comparing the dissipative and
elastic terms. It reads $\tau_c^\eta \sim \eta_l L_c^2 / C$ and
the corresponding minimal action for the tunneling process is
\begin{equation}
B_c^d \sim \tau_c^\eta L_c \frac{C u_c^2}{L_c^2} \sim \eta_l u_c^2 L_c.
\label{bdis}
\end{equation}

Now, we estimate the calculated values in order to better understand our
results. The friction coefficient per unit length $\eta_l = \eta / a$ can be estimated from the known data for
the metallic phase $(\nu = 0.1)$. By using the Drude formula, we can evalute
 $\eta = n e^2 / \sigma$ and the corresponding relaxation time $\tau = m / \eta$.
Here, $n$ is the hole concentration per unitary volume $V_0$, $n = \nu / V_0 \sim 10^{-1}/
200 \AA^3 = 0.5 \times 10^{21} $ cm$^{-3}$, 
 $\sigma$ is the normal state conductivity, $\sigma \sim 10^3 \, \Omega^{-1}$ 
cm$^{-1}$ \cite{Sun} and $m$ is the effective tight-binding mass of the carriers in
the CuO plane, $m/m_e \sim 1$ \cite{Math}, with $m_e$ denoting the free electron
mass. We then obtain $\eta \sim 10^{-13}$ g/s and $\tau \sim 10^{-14}$ s.
The relaxation time is assumed to be independent of the doping, and to have the samer
order of magnitude for the metallic and for the insulating states.  
The tunneling time $\tau_c \sim 10^{-12}$ s and $\tau_c^\eta \sim 10^{-10}$ s. This
indicates that the quantum dynamics is overdamped, and that a proper theory for 
describing the relaxation process should account in a more accurate way for the
dissipative term.

Let us now calculate the correction in the minimal action due to the presence
of the electrical field. In general, a quantum problem can be regarded as a
$(d + 1)$-dimensional generalization of a classical $d$-dimensional problem,
with the imaginary time being considered as an additional dimension. However,
one should notice that the disorder potential fluctuates in space, but not
in time. At high temperatures, the stripe jumps over the barrier and the time
needed for the jump is irrelevant. At low temperatures, instead, this time
is essential. Hence, for the classical motion the stripe can choose optimal
barriers [see Eq. (\ref{optbar})], whereas for tunneling the relevant barrier
scales like the average barrier $U_c (L/L_c)$ (see \cite{Gianni} for a more
detailed discussion). As a consequence, we obtain that in the limit of low
driving fields $E \ll E_c$, the minimal action reads
\begin{equation}
B \sim B_c \left( \frac{E_c}{E} \right)^{\mu_q},
\end{equation}
were $\mu_q = (1 + \zeta) / (2 - \zeta)$ and $B_c$ is $B_c^m$ or $B_c^d$
[given by Eqs. (\ref{bmas}) or (\ref{bdis}), respectively], depending
if we are considering the massive or the dissipative limits. Remembering
that in our case $\zeta = 2/3$, we finally obtain a quantum glassy exponent
$\mu_q = 5/4$. Near criticality, the results change considerably: as in the
thermal case, in this limit the quantum action exhibits a power law behavior,
\begin{equation}
B \sim B_c \left(1 - \frac{E}{E_c}\right)^\alpha,
\end{equation}
with $\alpha = 1$ for the massive stripe and $\alpha = 3/4$ for the
overdamped one. 

\section{DISCUSSIONS}

In this work we have succeeded in deducing a wave like equation for the motion
of a line of holes embedded in a antiferromagnetic phase, which we have chosen
to describe by a $t-J$ model. This was done through the application of
semi-classical methods to a fictious spin chain whose local spins can 
appropriately be related to the displacement difference between neighbouring
holes along a direction perpendicular to the line itself. Therefore we were
able to establish a connection between the microscopic parameters of the
$t-J$ model and the phenomenological mass and elastic coefficients of the
continuum theory by using the ficticious spin model as an intermediate step.

We have also extended the well-established ``collective pinning theory'', 
which has successfully been applied to vortices in superconductors and charge
density waves in quasi 1-D electronic systems, to the case of the wall of
holes. In so doing, we were able to estimate the threshold field for the
depinning of the wall as well as the energy barrier per unit length felt by
the trapped line. These quantities allowed us to compute the thermal and
quantal rates for the depinning of a single wall of holes. 

These results can be tested by measuring the conductance of a sample, which
presents the striped phase in its antiferromagnetic regime, as a function of
the external field (voltage). From our findings, we expect to have a
vanishingly small conductance up to the threshold field and then a gradual
tendency to recover the ohmic regime for $E > E_c$. This highly non-linear
behavior of the conductance is analogous to that observed in charge density
wave systems \cite{cdw}. It is also important to notice that in order to 
compare our results with the experimental ones, one should take into account
that inhomogeneities in the barrier height distribution may affect the
observed threshold field. Actually, the measured value is a lower bound for
the estimated voltage, and results differing by even one order of magnitude
from our findings would not be surprising. 

Finally, we would like to say some words about our approach to the damping
of the wall motion. We have chosen to use an entirely phenomenological
approach to the problem because it is not possible to describe dissipation from 
the present microscopic model. The first step in this direction would be to allow 
the spin system to respond to any change of the line configuration, i.e., we
should assume a finite stiffness for the magnetic system. This procedure is
planned to be done in a future publication. 

We are indebt with M.\ Bali\~ na, D.\ Baeriswyl, A.\ H.\ Castro Neto, and
H.\ Schmidt for fruitfull discussions. This work has been supported by the
DAAD-CAPES PROBRAL project number 415. 
N.\ H.\ acknowledges financial support from the Gottlieb Daimler- und Karl Benz-
Stiftung and the Graduiertenkolleg ``Physik nanostrukturierter Festk\"orper'', 
Universit\"at Hamburg.  Y.\ D.\ acknowledges financial support from the Otto
Benecke-Stiftung.

\newpage
\appendix

\section{}
Our aim here is to calculate the commutators on the right-hand side of Eq.\ 
(\ref{sigmanz}). With the help of the relation (\ref{com}), one obtains
$$
\hbar^2 \stackrel{..}{\sigma}_n^z = -\left[\left[\sigma_n^z,H\right],H\right] = 
$$
$$
=\frac{tJ}{\sigma^2}\Bigg\{
\left(\sigma _n^{+}\sigma _{n+1}^{-}\right)
\left(\mid \sigma_{n+1}^z \mid + \mid \sigma _n^z 
\mid - \mid \sigma _{n+1}^z - 1 \mid - \mid \sigma _n^z + 1 \mid \right)
$$
$$
- \left(\sigma _n^{-}\sigma _{n+1}^{+}\right)
\left(\mid \sigma_{n+1}^z \mid + \mid \sigma _n^z \mid - \mid \sigma _{n+1}^z 
+ 1 \mid - \mid \sigma _n^z - 1 \mid \right)
$$
$$
+\left(\sigma _n^{+}\sigma _{n-1}^{-}\right)
\left(\mid \sigma_{n-1}^z \mid + \mid \sigma _n^z 
\mid - \mid \sigma _{n-1}^z - 1 \mid - \mid \sigma _n^z + 1 \mid \right)
$$
$$
- \left(\sigma _n^{-}\sigma _{n-1}^{+}\right)
\left(\mid \sigma_{n-1}^z \mid + \mid \sigma _n^z \mid - \mid \sigma _{n-1}^z 
+ 1 \mid - \mid \sigma _n^z - 1 \mid \right) \Bigg\} + 
$$
$$
+\frac{2 t^2}{\sigma^4}
\Bigg\{\sigma_{n+1}^z 
\left(\sigma_n^{+}\sigma_n^{-}+\sigma_n^{-}\sigma_n^{+}+\sigma_n^{+}\sigma_{n+2}^{-}+
\sigma_n^{-}\sigma_{n+2}^{+}\right) 
$$
$$
+\sigma_{n-1}^z
\left(\sigma_n^{+}\sigma_n^{-}+\sigma_n^{-}\sigma_n^{+}+\sigma_n^{+}\sigma_{n-2}^{-}+
\sigma_n^{-}\sigma_{n-2}^{+}\right)\Big\}
$$
$$
-\sigma_n^z
\left(\sigma_{n+1}^{+}\sigma_{n+1}^{-}+\sigma_{n+1}^{-}\sigma _{n+1}^{+} + 
\sigma_{n-1}^{+}\sigma_{n-1}^{-}+\sigma_{n-1}^{-}\sigma_{n-1}^{+}
+ 2\sigma _{n-1}^{+}\sigma _{n+1}^{-}+2\sigma
_{n-1}^{-}\sigma _{n+1}^{+}\right)\Bigg\}
$$

Now, we consider the problem in the classical limit, when the operators $\sigma_n^z$
and $\sigma_n^\pm$ can be replaced by c-numbers. In spherical coordinates, we can
write
\begin{equation}
\sigma_n^z = \sigma \sin \alpha_n, \qquad \sigma_n^\pm = \sigma \exp (\pm i \phi_n
) \cos \alpha_n,
\end{equation}
where $\alpha_n = \pi / 2 - \theta_n$. Using that $\phi_n - \phi_{n -1} \sim 0$
and $\sigma_n^z \ll 1$, we obtain the classical equation of motion 
\begin{equation}
\hbar^2 \stackrel{..}{\alpha}_n = t \left(J + \frac{8 t}{\sigma^2} \right) 
\left(\alpha_{n+1} - 2 \alpha_n + \alpha_{n-1} \right). 
\end{equation}
Further, we take the limit $\sigma \to \infty$ and go to the continuous approximation.
In this limit, we can replace
\begin{equation}
\alpha_n \to \alpha (y), \qquad \alpha_{n \pm 1} \to \alpha(y) \pm a \alpha'(y) + 
\frac{a^2}{2} \alpha'' (y),
\end{equation}
and we eventually obtain the wave-like equation
\begin{equation}
\left(\frac{\partial^2}{\partial t'^2} - \frac{t Ja^2}{\hbar^2} 
\frac{\partial^2}{\partial y^2} \right) \alpha = 0.
\label{eqmalpha}
\end{equation}

\noindent{\bf FIGURE CAPTIONS}

{\bf Fig.\ 1:}
Regions corresponding to the diagonal and vertical striped phase obtained
from (a) exact diagonalization in small systems \cite{Prel-Zoto}; (b)
Hartree-Fock calculations \cite{Inui-Litt}.

{\bf Fig.\ 2:}
(a) Diagonal and (b) vertical stripe configurations.

{\bf Fig.\ 3:}
(a) Delocalized stripe (antiphase domain boundary); (b) Localized stripe
(hypothetical case corresponding to a phase domain boundary).

\newpage

\begin{figure}[h]

\begin{picture}(15,9)

\centerline{\epsfig{file=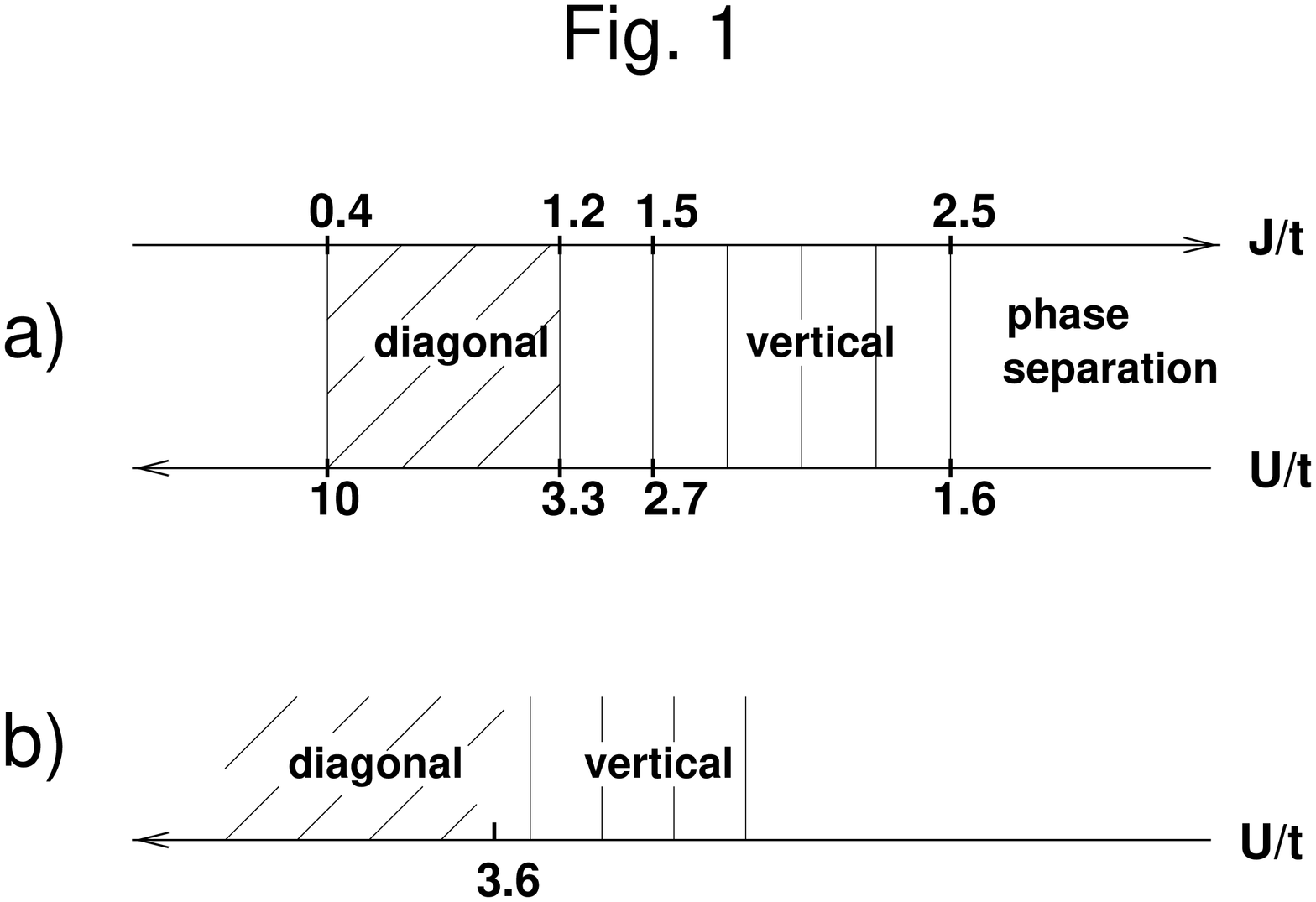,angle=-90,width=13cm}}

\end{picture}

\end{figure}

\vspace{11cm}

\begin{figure}[h]

\begin{picture}(15,9)

\centerline{\epsfig{file=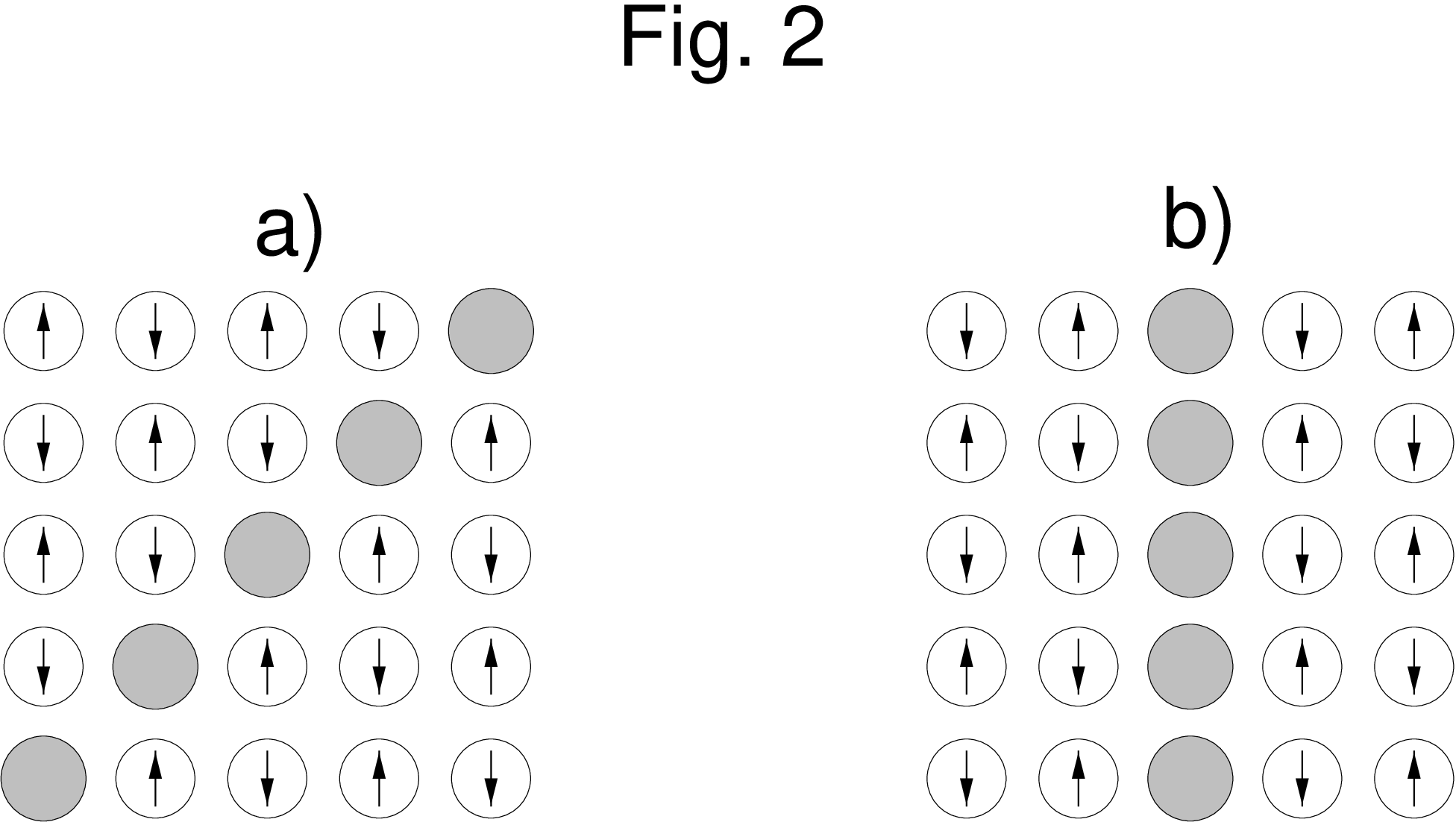,angle=-90,width=13cm}}

\end{picture}

\end{figure}

\newpage

\begin{figure}[h]

\begin{picture}(15,9)

\centerline{\epsfig{file=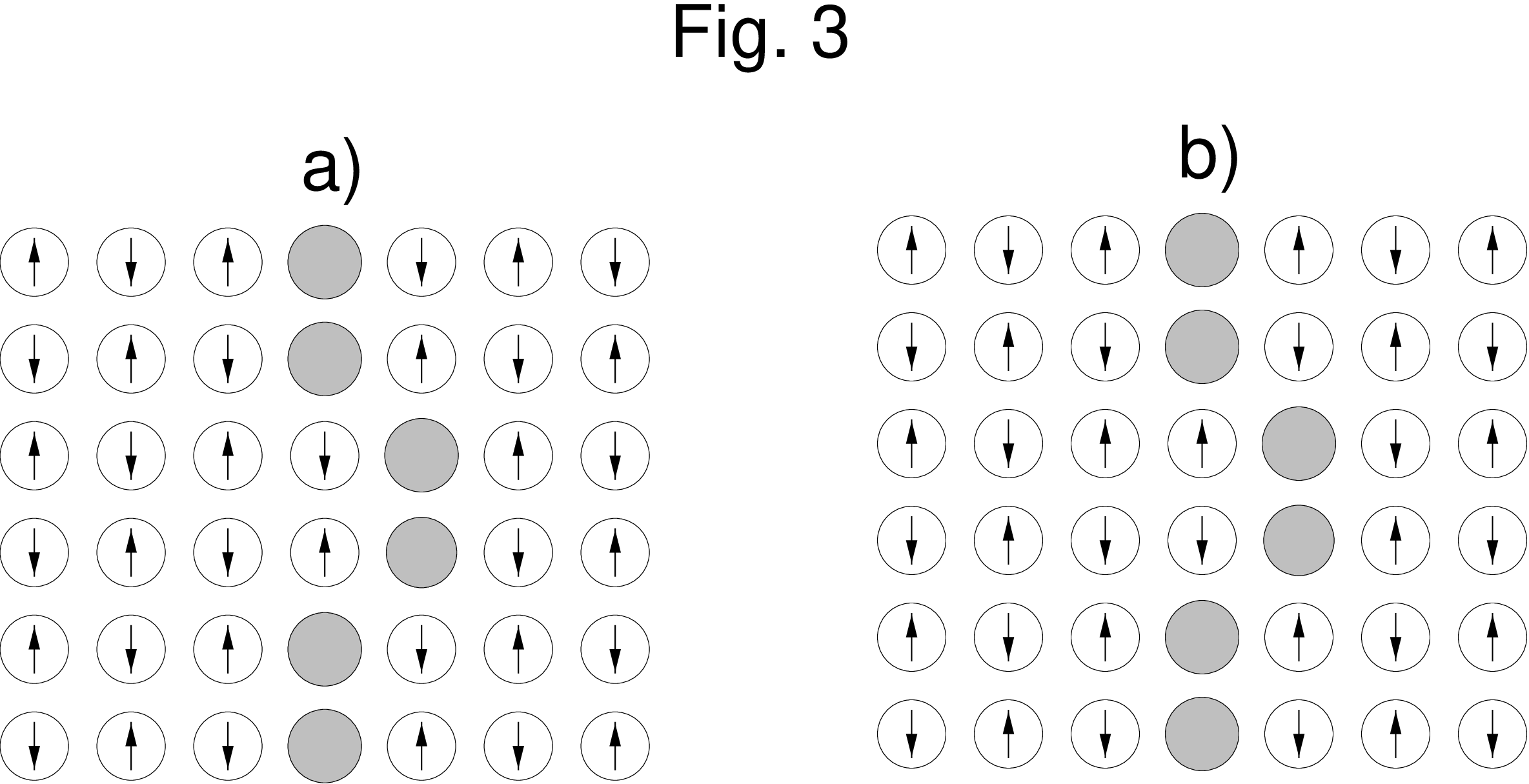,angle=-90,width=13cm}}

\end{picture}

\end{figure}

\end{document}